\documentclass[sigconf,screen]{acmart}

\usepackage{enumitem}
\usepackage{amsmath,amsfonts}
\usepackage{algorithmic}
\usepackage{graphicx}
\usepackage{textcomp}
\usepackage{xcolor}
\usepackage[english]{babel}
\usepackage{subcaption}
\usepackage{tabularx}
\usepackage{array}
\usepackage{url}

\usepackage{caption}
\usepackage{xspace}

\usepackage{lineno}
\usepackage{multirow}
\usepackage{setspace}

\usepackage{array}
\usepackage{tabularx}

\usepackage{xcolor}
\usepackage{pgfplots}
\usepackage{pgfplotstable}

\usepackage{tikzpagenodes}
\usepackage{eso-pic}

\usepackage{float}

\AtBeginDocument{%
  }

\copyrightyear{2025}
\acmYear{2025}
\setcopyright{acmlicensed}\acmConference[ISCA '25]{Proceedings of the 52nd Annual International Symposium on Computer Architecture}{June 21--25, 2025}{Tokyo, Japan}
\acmBooktitle{Proceedings of the 52nd Annual International Symposium on Computer Architecture (ISCA '25), June 21--25, 2025, Tokyo, Japan}
\acmDOI{10.1145/3695053.3731412}
\acmISBN{979-8-4007-1261-6/2025/06}

\newcommand{\dsvii}{DeepSeek-V2}

\newcommand{\dsviii}{DeepSeek-V3}

\begin{document}

\title{Insights into DeepSeek-V3: Scaling Challenges and Reflections on Hardware for AI Architectures}

\newcommand{\MyAuthors}{
Chenggang Zhao,
Chengqi Deng,
Chong Ruan,
Damai Dai,
Huazuo Gao,
Jiashi Li,
Liyue Zhang,
Panpan Huang,
Shangyan Zhou,
Shirong Ma,
Wenfeng Liang,
Ying He,
Yuqing Wang,
Yuxuan Liu,
Y.X. Wei
}

\author{Chenggang Zhao}
\affiliation{
  \institution{DeepSeek-AI}
  \city{Beijing}
  \country{China}
}
\email{chenggangz@deepseek.com}

\author{Chengqi Deng}
\affiliation{
  \institution{DeepSeek-AI}
  \city{Beijing}
  \country{China}
}
\email{cq.deng@deepseek.com}

\author{Chong Ruan}
\affiliation{
  \institution{DeepSeek-AI}
  \city{Beijing}
  \country{China}
}
\email{chong.ruan@deepseek.com}

\author{Damai Dai}
\affiliation{
  \institution{DeepSeek-AI}
  \city{Beijing}
  \country{China}
}
\email{damai.dai@deepseek.com}

\author{Huazuo Gao}
\affiliation{
  \institution{DeepSeek-AI}
  \city{Beijing}
  \country{China}
}
\email{gaohuazuo@deepseek.com}

\author{Jiashi Li}
\affiliation{
  \institution{DeepSeek-AI}
  \city{Beijing}
  \country{China}
}
\email{js.li@deepseek.com}

\author{Liyue Zhang}
\authornote{Yuqing Wang and Liyue Zhang are the corresponding authors of this paper. Authors are listed in alphabetical order of their first names.}
\affiliation{
  \institution{DeepSeek-AI}
  \city{Beijing}
  \country{China}
}
\email{ly.zhang@deepseek.com}

\author{Panpan Huang}
\affiliation{
  \institution{DeepSeek-AI}
  \city{Beijing}
  \country{China}
}
\email{pp.huang@deepseek.com}

\author{Shangyan Zhou}
\affiliation{
  \institution{DeepSeek-AI}
  \city{Beijing}
  \country{China}
}
\email{sy.zhou@deepseek.com}

\author{Shirong Ma}
\affiliation{
  \institution{DeepSeek-AI}
  \city{Beijing}
  \country{China}
}
\email{mashirong.2000@deepseek.com}

\author{Wenfeng Liang}
\affiliation{
  \institution{DeepSeek-AI}
  \city{Beijing}
  \country{China}
}
\email{wenfeng.liang@deepseek.com}

\author{Ying He}
\affiliation{
  \institution{DeepSeek-AI}
  \city{Beijing}
  \country{China}
}
\email{ying.he@deepseek.com}

\author{Yuqing Wang}
\authornotemark[1]
\affiliation{
  \institution{DeepSeek-AI}
  \city{Beijing}
  \country{China}
}
\email{wangyq@deepseek.com}

\author{Yuxuan Liu}
\affiliation{
  \institution{DeepSeek-AI}
  \city{Beijing}
  \country{China}
}
\email{liuyuxuan@deepseek.com}

\author{Y.X. Wei}
\affiliation{
  \institution{DeepSeek-AI}
  \city{Beijing}
  \country{China}
}
\email{weiyx@deepseek.com}



\begin{abstract}
The rapid scaling of large language models (LLMs) has unveiled critical limitations in current hardware architectures, including constraints in memory capacity, computational efficiency, and interconnection bandwidth. DeepSeek-V3, trained on 2,048 NVIDIA H800 GPUs, demonstrates how hardware-aware model co-design can effectively address these challenges, enabling cost-efficient training and inference at scale. This paper presents an in-depth analysis of the DeepSeek-V3/R1 model architecture and its AI infrastructure, highlighting key innovations such as Multi-head Latent Attention (MLA) for enhanced memory efficiency, Mixture of Experts (MoE) architectures for optimized computation-communication trade-offs, FP8 mixed-precision training to unlock the full potential of hardware capabilities, and a Multi-Plane Network Topology to minimize cluster-level network overhead. Building on the hardware bottlenecks encountered during DeepSeek-V3's development, we engage in a broader discussion with academic and industry peers on potential future hardware directions, including precise low-precision computation units, scale-up and scale-out convergence, and innovations in low-latency communication fabrics. These insights underscore the critical role of hardware and model co-design in meeting the
escalating demands of AI workloads, offering a practical blueprint for innovation in next-generation AI systems.
\end{abstract}




\maketitle

\AddToShipoutPictureBG*{%
  \AtPageLowerLeft{%
    \raisebox{1.5cm}[0pt][0pt]{%
      \hspace{2.25cm}%
      \fbox{%
        \begin{minipage}{0.95\textwidth}
          \footnotesize
          This is the author's version of the work. It is posted here for your personal use. Not for redistribution.\\
          The definitive version appeared as part of the Industry Track in \textit{Proceedings of the 52nd Annual International Symposium on Computer Architecture (ISCA '25)}.
        \end{minipage}
      }
    }
  }
}

\section{Introduction}

\subsection{Background}
Large Language Models~(LLMs) have undergone rapid evolution in recent years, driven by iterative advancements in model design, computational power, and data availability. In 2024, groundbreaking models such as GPT4o~\citep{gpt4o}, LLaMa-3~\citep{llama3}, Claude 3.5 Sonnet~\citep{claude35sonnet},  Grok-2~\citep{grok2},  Qwen2.5~\cite{qwen2.5}, Gemini-2~\citep{gemini2} and our DeepSeek-V3~\citep{dsviii} have showcased remarkable progress, further narrowing the gap towards Artificial General Intelligence~(AGI). As the Scaling Laws~\cite{scalinglaw} shows, increasing model size, training data, and computational resources leads to substantial improvements in model performance, underscoring the pivotal role of scaling in advancing AI capabilities. Collectively, these developments have ushered in an era where scaling model size and computational power is seen as the key to unlocking higher levels of intelligence.\let\thefootnote\relax

Recent developments, reasoning models such as OpenAI's o1/o3 series models~\citep{openai_o1,openaio3}, DeepSeek-R1~\citep{dsr1}, Claude-3.7 Sonnet~\citep{claude37sonnet}, Gemini 2.5 Pro~\citep{gemini2_5}, Seed1.5-Thinking~\citep{seed2025seed15thinkingadvancingsuperbreasoning} and Qwen3~\citep{qwen3} have demonstrated not only the benefits conferred by large-scale architectures, but also the necessity of improving inference efficiency, particularly in handling longer contexts and achieving greater reasoning depth. These advancements underscore the need for faster and more efficient inference, consequently placing ever-increasing demands on computational resources.

To meet these challenges, industry leaders such as Alibaba, ByteDance, Google, xAI and Meta have deployed colossal training clusters~\cite{10.1145/3579371.3589350,mudigere_software-hardware_2023-1,10.1145/3651890.3672233,jiang_megascale_2024,10.1145/3651890.3672265,xaicolossus}, featuring tens or even hundreds of thousands of GPUs or TPUs. While such massive infrastructures have enabled the development of state-of-the-art models, their exorbitant costs present significant barriers for smaller research teams and organizations. Despite these barriers, open-source startups such as DeepSeek~\cite{dsvi,dsvii,dsviii,dscodervii,dsr1} and Mistral~\cite{mistral,mixtral8x22b} are also striving to develop state-of-the-art models.
Among them, DeepSeek has especially demonstrated that effective software-hardware co-design can enable cost-efficient training of large models, leveling the playing field for smaller teams.

Building on this tradition, DeepSeek-V3~\cite{dsviii} represents a new milestone in cost-effective training. By leveraging just 2,048 NVIDIA H800 GPUs, DeepSeek-V3 achieves state-of-the-art performance. This achievement aligns with the commitment to advance AI through practical and scalable solutions, as previously demonstrated in the cost-effective architecture of Fire-Flyer AI-HPC~\cite{10793193}. The practices and insights derived from DeepSeek-V3 demonstrate how existing hardware resources can be harnessed to their fullest potential, offering valuable lessons for the broader AI and HPC communities.

\subsection{Objectives}
This paper does not aim to reiterate the detailed architectural and algorithmic specifics of DeepSeek-V3, which are extensively documented in its technical report~\cite{dsviii}. Instead, it adopts a dual perspective—spanning hardware architecture and model design—to explore the intricate interplay between them in achieving cost-efficient large-scale training and inference. By examining this synergy, we aim to provide actionable insights for scaling LLMs efficiently without sacrificing performance or accessibility.

Specifically, the paper focuses on:
\begin{itemize}[leftmargin=12pt,topsep=0pt]
    \item \textbf{Hardware-Driven Model Design:} Analyze how hardware features, such as FP8 low-precision computation and scale-up/scale-out network properties, informed the architectural choices in DeepSeek-V3.
    \item \textbf{Mutual Dependencies Between Hardware and Models:} Investigate how hardware capabilities shape model innovation and how the evolving demands of LLMs drive the need for next-generation hardware.
    \item \textbf{Future Directions for Hardware Development:} Derive actionable insights from DeepSeek-V3 to guide the co-design of future hardware and model architectures, paving the way for scalable, cost-efficient AI systems.
\end{itemize}
\subsection{Structure of this Paper}
The remainder of this paper is organized as follows. Section~\ref{sec:design_principles}  explores the design principles underpinning DeepSeek-V3 model architecture, highlighting key innovations such as Multi-head Latent Attention, Mixture-of-Experts optimizations and Multi-Token Prediction Module. Section~\ref{sec:hardware_features} illustrates how our model architecture pursues low-precision computation and communication. Section~\ref{sec:scale_up_optimizations} includes scale-up interconnection optimizations, discusses scale-up/scale-out convergence, and explores how hardware features influence parallelism and expert selection strategies. Section~\ref{sec:scale_out_optimizations} focuses on scale-out network optimizations, including multi-plane network co-designs and low-latency interconnects. Besides current limitations and future suggestions mentioned in Section~\ref{sec:hardware_features}$\sim$\ref{sec:scale_out_optimizations}, Section~\ref{sec:future_hardware} elaborates on more critical insights from DeepSeek-V3, and identifies directions for future hardware and model co-design.

\begin{figure*}[!t]
\centering
\includegraphics[width=0.99\linewidth]{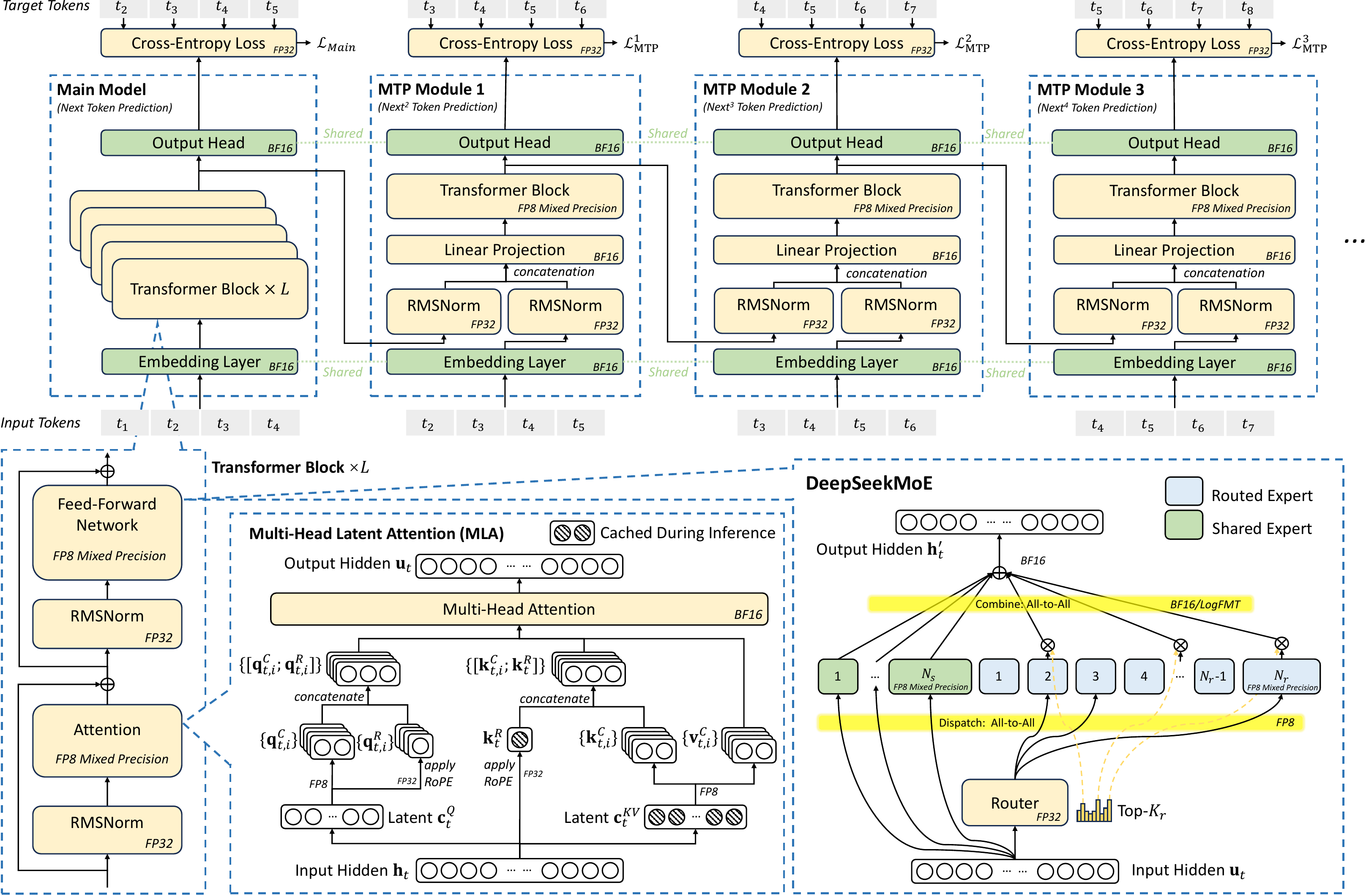}
\vspace{-5pt} 
\caption{
    Basic architecture of DeepSeek-V3. Built upon DeepSeek-V2’s MLA and DeepSeekMoE, a Multi-Token Prediction Module and FP8 mixed-precision training are introduced to enhance inference and training efficiency. The figure indicates the precision used for computations in different parts of the architecture. All components take inputs and outputs in BF16.
}
\label{fig:basic_arch}
\vspace{-5pt}
\end{figure*}

\section{Design Principles for DeepSeek Models}
\label{sec:design_principles}

The development of \textbf{DeepSeek-V3} exemplifies a hardware-aware approach to scaling LLMs, where each design decision was carefully aligned with hardware constraints to optimize performance and cost efficiency.

As shown in Figure~\ref{fig:basic_arch}, \dsviii{} employs the
\textbf{DeepSeekMoE}~\cite{deepseekmoe} and \textbf{Multi-head Latent Attention (MLA)}~\cite{dsvii} architectures that have been proven effective in \dsvii{}~\cite{dsvii}. 
DeepSeekMoE unlocks the potential of MoE architecture, while MLA drastically reduces memory consumption by compressing Key-Value (KV) caches. 
In addition, \textbf{DeepSeek-V3} incorporates \textbf{FP8 mixed-precision training}, significantly lowering computational costs and making large-scale training more practical without compromising model quality. To improve the inference speed, DeepSeek-V3 integrates speculative decoding based on its \textbf{Multi-Token Prediction Module}, which significantly increases the generation speed. Beyond model architecture, we also explored cost-efficient AI infrastructure by deploying a \textbf{Multi-Plane} two-layer Fat-Tree network to replace a traditional three-layer Fat-Tree topology, reducing cluster networking costs.

These innovations aim to address three core challenges in scaling LLMs—\textbf{memory efficiency}, \textbf{cost-effectiveness}, and \textbf{inference speed}—which are explored in detail in the following subsections.

\subsection{Memory Efficiency}

LLMs generally require significant memory resources, with memory demands increasing by more than 1000\% per year. In contrast, the growth rate of high-speed memory (e.g., HBM) capacity is much slower, typically less than 50\% per year~\citep{10477550}. While multi-node parallelism is a viable solution to address memory limitations, optimizing memory usage at the source remains a crucial and effective strategy.

\vspace{-0.3em}
\subsubsection{Low-Precision Models}
Compared to models that utilize BF16 for weights, FP8 significantly reduces memory consumption by half, effectively alleviating the AI memory wall challenge. A detailed discussion of low-precision techniques is provided in Section~\ref{sec:hardware_features} Low-Precision Driven Design.

\vspace{-0.3em}
\subsubsection{Reducing KV Cache with MLA}

For LLM inference, user requests often involve multi-turn conversations. To handle these efficiently, the context from previous requests is cached in what is commonly referred to as the \textit{KV cache}. KV cache addresses this challenge by caching the \textbf{Key} and \textbf{Value} vectors of previously processed tokens, eliminating the need to recompute them for subsequent tokens. During each inference step, the model only computes the Key and Value vectors for the current token and performs attention computation by combining them with the cached Key-Value pairs from the history. This incremental computation reduces the complexity of generating each token to \(O(N)\), making it efficient when processing long sequences or multi-turn inputs. However, it introduces a memory-bound bottleneck because the computation shifts from GEMM to GEMV, which has a much lower compute-to-memory ratio. With modern hardware offering hundreds of TFLOPS, GEMV quickly becomes limited by memory bandwidth, making memory access the primary bottleneck.

To address this bottleneck, we employ \textbf{Multi-head Latent Attention (MLA)}~\cite{dsvii} that compresses the KV representations of all attention heads into a smaller latent vector using a projection matrix, which is jointly trained with the model. During inference, only the latent vector needs to be cached, significantly reducing memory consumption compared to storing the KV cache for all attention heads.

In addition to MLA, several other approaches have been proposed to reduce the size of the KV cache. These methods are highly valuable and provide significant inspiration for advancements in memory-efficient attention mechanisms:

\begin{itemize}[leftmargin=12pt,topsep=0pt]
    \item \textbf{Shared KV (Grouped-Query Attention, GQA; Multi-Query Attention, MQA):} Instead of maintaining separate KV pairs for each attention head, multiple heads share a single set of KV pairs, significantly compressing KV storage. Representative methods include GQA~\citep{ainslie2023gqa} and MQA~\citep{mqa}.
    \item \textbf{Windowed KV:} For long sequences, only a sliding window of KV pairs is retained in the cache, discarding results outside the window. While this reduces storage, it compromises long-context reasoning. Representative methods include Longformer~\cite{Beltagy2020Longformer} and related architectures.
    \item \textbf{Quantized Compression:}  KV pairs are stored using low-bit representations~\citep{hooper2024kvquant,liu2024kivi,kang2024gear}, further reducing memory usage. Quantization achieves significant compression with minimal impact on model performance.
\end{itemize}

\begin{table}[!t]
\centering
\small
\caption{KV cache size comparison (BF16 precision): DeepSeek-V3 (MLA) largely reduces KV cache size compared to other models using GQA.}
\vspace{-5pt}
\label{tab:kvcache_comparison}
\begin{tabular}{|l|c|c|}
\hline
\textbf{Model}          & \textbf{KV Cache Per Token} & \textbf{Multiplier} \\ \hline
\textbf{DeepSeek-V3 (MLA)}    & ~~70.272 KB                         & 1x                                    \\ \hline
\textbf{Qwen-2.5 72B (GQA)}    & 327.680 KB                       & 4.66x                                  \\ \hline
\textbf{LLaMA-3.1 405B (GQA)}     & 516.096 KB                          & 7.28x    \\ \hline
\end{tabular}
\vspace{-10pt}
\end{table}

Table~\ref{tab:kvcache_comparison} compares the KV cache memory usage per token among DeepSeek-V3, Qwen-2.5 72B~\citep{qwen2.5}, and LLaMA-3.1 405B~\citep{llama3_1_405b}. By adopting MLA, DeepSeek-V3 achieves a significant reduction in KV cache size, requiring only 70 KB per token, substantially less than LLaMA-3.1 405B's 516 KB and Qwen-2.5 72B's 327 KB.
This reduction highlights the efficiency of MLA in compressing KV representations compared to GQA-based methods. The ability to achieve such a significant reduction in memory consumption makes DeepSeek-V3 particularly well-suited for scenarios involving long-context processing and resource-constrained environments, enabling more scalable and cost-effective inference.

\subsubsection{Future Directions and Perspectives on Resource-Efficient Techniques}
\label{section:future_model_techs}
While reducing the size of the KV cache is a promising method for improving memory efficiency, the quadratic complexity inherent in Transformer-based autoregressive decoding remains a formidable challenge, especially for extremely long contexts. Recent research efforts, such as Mamba-2~\citep{10.5555/3692070.3692469} and Lightning Attention\citep{10.5555/3692070.3693758}, investigate linear-time alternatives that offer new possibilities for balancing computational cost and model performance.  In addition, approaches such as sparse attention~\cite{dsnsa}, which seek to compress and sparsely activate attention keys and values, represent another attempt at overcoming the computational challenges associated with attention.  We look forward to collaborative progress with the broader community toward breakthroughs in this area.

\subsection{Cost-Effectiveness of MoE Models}

For sparse computing, we have developed DeepSeekMoE, an advanced \textbf{Mixture of Experts (MoE)} architecture, which is illustrated in the lower right part of Figure~\ref{fig:basic_arch}.
The advantages of MoE models lie in two folds. 

\subsubsection{Reducing Computational Requirements for Training}

The primary advantage of the MoE architecture lies in its ability to significantly reduce training costs. By selectively activating only a subset of expert parameters, MoE models allow the total parameter count to scale up dramatically while keeping computational requirements modest. For example, \textbf{DeepSeek-V2} features 236B parameters, but only 21B parameters are activated per token. Similarly, \textbf{DeepSeek-V3} expands to 671B parameters—nearly three times the size of V2—while keeping the activation per token at just 37B. In comparison, dense models such as Qwen2.5-72B and LLaMa3.1-405B require all parameters to be active during training. 

As shown in Table~\ref{tab:moe_computation}, the total computational cost for DeepSeek-V3 is approximately 250 GFLOPS per token, whereas the 72B dense model requires 394 GFLOPS and the 405B dense model requires 2448 GFLOPS. This demonstrates that MoE models achieve comparable or even superior performance to dense models while consuming an order of magnitude less computational resources.

\subsubsection{Advantages for Personal Use and On-Premises Deployment}

In a future where personalized LLM agents~\citep{agentsurvey2025} become ubiquitous, MoE models offer unique advantages in single-request scenarios. Because only a subset of parameters is activated per request, memory and computational demands are greatly reduced. For example, \textbf{DeepSeek-V2} (236B parameters) activates just 21B parameters during inference. 
This enables PCs with AI SoC chips~\citep{apple_m4,dgx_spark,amd_395} to achieve nearly 20 tokens per second (TPS), or even twice that speed, which is more than sufficient for personal use. In contrast, dense models of similar capability (e.g., 70B parameters) typically reach only single-digit TPS on similar hardware. 

Notably, the increasingly popular KTransformers~\citep{ktransformers} inference engine allows the complete DeepSeek-V3 model to run on a low-cost server equipped with a consumer GPU (costing approximately \$10,000), while still achieving nearly 20 TPS.

This efficiency makes MoE architectures suitable for local deployments and single-user scenarios, where hardware resources are often limited. By minimizing memory and computational overhead, MoE models can deliver high-quality inference performance without requiring expensive infrastructure.

\begin{table}[!t]
\centering
\small
\caption{Comparison of computational costs for training MoE and dense models: Computational cost per token is measured, assuming a sequence length of 4096.}
\label{tab:moe_computation}
\vspace{-5pt}
\begin{tabular}{|l|c|c|}
\hline
\textbf{Model}          & \textbf{Size} & \textbf{Training Cost} \\ \hline
\textbf{DeepSeek-V2 MoE}    & 236B                      & 155 GFLOPS/Token                     \\ \hline
\textbf{DeepSeek-V3 MoE}    & 671B                      & 250 GFLOPS/Token                     \\ \hline
\textbf{Qwen-72B Dense}            & 72B                       & 394 GFLOPS/Token                     \\ \hline
\textbf{LLaMa-405B Dense}           & 405B                      & 2448 GFLOPS/Token                    \\ \hline
\end{tabular}
\vspace{-5pt}
\end{table}

\subsection{Increasing Inference Speed}

\subsubsection{Overlapping Computation and Communication: Maximizing Throughput}
\label{sec:maximizing_throughput}
Inference speed encompasses both system-wide maximum throughput and single-request latency. To maximize throughput, our model is architected from the outset to leverage dual micro-batch overlap~\cite{dsv3_profile_data,deepep2025}, intentionally overlapping communication latency with computation. As demonstrated in our online inference system and supported by open-source profiling data~\cite{dsv3_profile_data}, we decouple the computation of MLA and MoE into two distinct stages. While one micro-batch executes a portion of MLA or MoE computation, the other micro-batch simultaneously performs the corresponding dispatch communication. Conversely, during the computation phase of the second micro-batch, the first micro-batch undergoes the combine communication step. This pipelined approach enables seamless overlap of all-to-all communication with ongoing computation, ensuring that the GPU remains fully utilized at all times. Moreover, in production, we adopt a prefill and decode disaggregation architecture~\citep{298687}, assigning large batch size prefill and latency-sensitive decode requests to different expert parallelism group sizes. This strategy ultimately maximizes system throughput under real-world service conditions.

\subsubsection{Inference Speed Limits}
\label{section:inference_speed_limits}
This section focuses on the decode output speed of LLM services, typically measured in \textbf{Time Per Output Token (TPOT)}. TPOT is a critical metric for user experience, and it also directly impacts the responsiveness of reasoning models such as OpenAI's o1/o3 and DeepSeek-R1, which rely on the inference length to enhance their intelligence.

For MoE models, achieving high inference speed relies on efficiently deploying expert parameters across computing devices. To achieve the fastest possible inference speed, each device should ideally perform computations for a single expert (or multiple devices should collaboratively compute a single expert if necessary). However, \textbf{Expert Parallelism (EP)} requires routing tokens to the appropriate devices, which involves \texttt{all-to-all} communication across the network. As a result, the upper limit of MoE inference speed is dictated by interconnection bandwidth.

Consider a system where each device holds one expert's parameters and processes approximately 32 tokens at a time. This token count strikes a balance between compute-to-memory ratio and communication latency. And this token count ensures that each device processes an equal batch size during expert parallelism, allowing the communication time to be easily calculated.

For a system interconnected with CX7 400Gbps InfiniBand (IB) NICs, the time required for the two \texttt{all-to-all} communications in EP is calculated as follows:
\[
\text{Comm. Time} = (1\text{Byte} + 2\text{Bytes}) \times 32 \times 9 \times 7\text{K} / 50\text{GB/s} = 120.96\mu s
\]
Here, dispatch uses FP8 (1 byte), while combine uses BF16 (2 bytes), and the hidden size of each token is approximately 7K. The factor 9 indicates that each token is transferred to 8 routed experts and 1 shared expert. Network latency is not included in this calculation.

As discussed in Section~\ref{sec:maximizing_throughput}, maximizing throughput necessitates the use of dual micro-batch overlap. In this strategy, our theoretical best-case analysis assumes that computation overhead is minimized, so the upper bound on performance is determined by communication latency. In practical inference workloads, however, request contexts are often much longer, and MLA computations typically dominate execution time. Thus, this analysis represents an idealized scenario under dual micro-batch overlap. Under this assumption, the total time per layer can be formulated as:

\vspace{-5pt}
\[
\text{Total Time Per Layer} = 2 \times 120.96\mu s = 241.92\mu s
\]
With 61 layers in DeepSeek-V3, the total inference time is:
\[
\text{Total Inference Time} = 61 \times 241.92\mu s = 14.76\text{ms}
\]
Thus, the theoretical upper limit for this system is approximately \textbf{14.76 ms TPOT}, equivalent to  \textbf{67 tokens per second}. However, in practice, factors such as communication overhead, latency, incomplete bandwidth utilization, and computational inefficiencies reduce this number.

By contrast, if a high-bandwidth interconnect like GB200 NVL72 (900GB/s unidirectional bandwidth across 72 GPUs) were used, the communication time per EP step drops to:
\[
\text{Comm. Time} = (1\text{Byte} + 2\text{Bytes}) \times 32 \times 9 \times 7\text{K} / 900\text{GB/s} = 6.72\mu s
\]
Assuming perfect overlap between computation and communication, this would yield a theoretical upper limit of \textbf{0.82 ms TPOT}, or approximately \textbf{1200 tokens per second}. However, this figure is purely theoretical and does not account for the substantial drop in GPU efficiency at small batch sizes; in real deployments, actual throughput will be significantly lower. Nonetheless, this calculation vividly illustrates the transformative potential of high-bandwidth scale-up networks in accelerating large-scale model inference.

While MoE models exhibit good scalability, achieving high inference speeds by increasing hardware resources alone is cost-prohibitive. 
Therefore, software and algorithms must also contribute to improving inference efficiency.

\subsubsection{Multi-Token Prediction}

Inspired by ~\citet{DBLP:conf/icml/GloeckleIRLS24}, DeepSeek-V3 introduces a \textbf{Multi-Token Prediction (MTP)} framework, which simultaneously enhances model performance and improves inference speed.
During inference, traditional autoregressive models generate one token at a decoding step, leading to sequential bottlenecks.
MTP mitigates this issue by enabling the model to generate additional candidate tokens at a lower cost and verify them in parallel, similar to previous self-drafting-based speculative decoding approaches~\cite{DBLP:conf/icml/CaiLGPLCD24, DBLP:conf/icml/LiW0024}. This framework significantly accelerates inference without compromising accuracy.

As illustrated in the top part of Figure~\ref{fig:basic_arch}, each MTP module uses a single layer, which is much more lightweight than the full model, to predict additional tokens, enabling parallel verification of multiple candidate tokens. 
Although slightly hurting the throughput, this approach significantly improves the end-to-end generation latency. 
The real world practice data demonstrates that an MTP module achieves an acceptance rate of 80\% to 90\% for predicting the second subsequent token, which increases the generation TPS by 1.8x compared to the scenario without the MTP module.

Moreover, by predicting multiple tokens per step, MTP increases the inference batch size, which is crucial for boosting EP computational intensity and hardware utilization. Such algorithmic innovations are vital for fast and cost-effective inference in DeepSeek-V3.

\subsubsection{High Inference Speed for Reasoning Models and Test-Time Scaling}
Test-time scaling in LLMs, exemplified by OpenAI's o1/o3 series~\cite{openai_o1,openaio3}, has enabled significant advances in mathematical reasoning, programming, and general reasoning by dynamically adjusting computational resources during inference. Subsequent models—including DeepSeek-R1~\citep{dsr1}, Claude-3.7 Sonnet~\citep{claude37sonnet}, Gemini 2.5 Pro~\citep{gemini2_5}, Seed1.5-Thinking~\citep{seed2025seed15thinkingadvancingsuperbreasoning}, and Qwen3~\citep{qwen3}—have adopted similar strategies and achieved notable improvements in these tasks.

For these reasoning models, high token output speed is of paramount importance. In reinforcement learning (RL) workflows—such as PPO~\citep{schulman2017proximalpolicyoptimizationalgorithms}, DPO~\citep{rafailov2024directpreferenceoptimizationlanguage} and GRPO~\citep{shao2024deepseekmathpushinglimitsmathematical}—the necessity to rapidly generate large numbers of samples makes inference throughput a critical bottleneck. Likewise, prolonged reasoning sequences can increase user wait times, reducing the practical usability of such models. As a result, optimizing inference speed through synergistic hardware and software innovations is indispensable for advancing the efficiency of reasoning models. However, effective strategies for accelerating inference and expediting RL training remain active areas of investigation, as discussed in Section~\ref{section:future_model_techs}. We encourage the broader community to collaboratively explore and develop novel solutions to these ongoing challenges.

\subsection{Technique Validation Methodology}

Each acceleration technique undergoes rigorous empirical validation to evaluate its accuracy impact, including MLA, FP8 mixed-precision computation, and network co-designed MoE gate routing. Given the prohibitive cost of exhaustive ablation on full-scale models, we adopt a hierarchical and resource-efficient validation pipeline. Each technique is first validated extensively on small-scale models, followed by minimal large-scale tuning, and finally integrated in a single, comprehensive training run.
For instance, we first conducted fine-grained FP8 training ablation studies on both 16B and 230B DeepSeek-V2 models before final integration. Under these controlled settings, the relative accuracy loss compared to BF16 remains below 0.25\%, attributable to our use of high-precision accumulation and fine-grained quantization strategies.

\vspace{-0.2em}
\section{Low-Precision Driven Design}
\label{sec:hardware_features}
\subsection{FP8 Mix-Precision Training}
Quantization techniques such as GPTQ~\cite{frantar2022gptq} and AWQ~\cite{lin2023awq} have been widely used to reduce bit-widths to 8-bit, 4-bit, or even lower, significantly reducing memory requirements. However, these techniques are primarily applied during inference to save memory, rather than in the training phase. NVIDIA's Transformer Engine has supported FP8 mixed-precision training for some time, but prior to DeepSeek-V3, there were no open-source large models leveraging FP8 for training. Through deep collaboration between our infrastructure and algorithm teams, and after extensive experimentation and innovation, we developed an FP8-compatible training framework for MoE models. Figure~\ref{fig:basic_arch} shows the computational components where FP8-precision forward and backward processes are utilized in the training pipeline. Fine-grained quantization is applied, i.e., tile-wise 1x128 quantization for activations and block-wise 128x128 quantization for model weights. Further technical details of our FP8 framework are documented in the DeepSeek-V3 technical report~\citep{dsviii}, and our fine-grained FP8 GEMM implementation has been open-sourced in DeepGEMM~\citep{deepgemm2025}.

\vspace{-0.2em}
\subsubsection{Limitations:}
While FP8 has great potential for accelerating training, several hardware limitations need to be addressed to fully exploit its capabilities:
\begin{itemize}[leftmargin=10pt]
    \item \textbf{FP8 Accumulation Precision:} FP8 uses constrained accumulation precision in Tensor Cores, affecting the stability for training large models, particularly on NVIDIA Hopper GPUs. After aligning 32 mantissa products by right-shifting based on the maximum exponent, the Tensor Core only maintains their highest 13 fraction bits for addition, and truncates bits exceeding this range. Addition results are accumulated to FP22 registers (1 sign bit, 8 exponent bits, and 13 mantissa bits)\citep{zhang2025sageattention2efficientattentionthorough}. The term ``FP22'' follows the naming used in the cited work, and the FP8 precision issue on Hopper GPUs had also been observed across industry by early 2024.
    \item \textbf{Fine-Grained Quantization Challenges:} Fine-grained quantization such as tile-wise and block-wise quantization introduces large dequantization overhead in transporting the partial results from Tensor Cores to CUDA Cores for scaling factor multiplication. This incurs frequent data movements, reducing computational efficiency and complicating hardware utilization.
\end{itemize}

\vspace{-0.2em}
\subsubsection{Suggestions:}
To address the limitations of existing hardware, we have the following suggestions for future designs:
\begin{itemize}[leftmargin=10pt]
    \item \textbf{Increased Accumulation Precision:} Hardware should improve the accumulation register precision to an appropriate value (e.g. FP32), or support a configurable accumulation precision, enabling a trade-off between performance and accuracy for different requirements of training and inference in various models.
    \item \textbf{Native Support for Fine-Grained Quantization:} Hardware should natively support fine-grained quantization, enabling Tensor Cores to receive scaling factors and implement matrix multiplication with group scaling. In this way, the whole partial sum accumulation and dequantization can be completed directly inside Tensor Cores until the final result is produced, avoiding frequent data movements to reduce dequantization overhead. A notable industrial implementation of this approach is NVIDIA Blackwell's support for \textbf{microscaling data format}~\citep{rouhani2023microscalingdataformatsdeep}, which exemplifies the practical benefits of native quantization at scale.
\end{itemize}

\vspace{-0.4em}
\subsection{LogFMT: Communication Compression}
\label{sec:logfmt}

In the current DeepSeek-V3 architecture, we employ low-precision compression for network communication. During EP parallelism, tokens are dispatched using fine-grained FP8 quantization, reducing communication volume by 50\% compared to BF16. This significantly lowers communication time. While the \texttt{combine} stage still uses higher precision (e.g., BF16) due to accuracy requirements, we are actively testing FP8, custom precision formats (e.g., E5M6) and mixing FP8-BF16 for further reductions.

Besides these traditional floating point formats, we also tried a new data type, named \textbf{Logarithmic Floating-Point Formats (LogFMT-nBit)}, where $n$ is the number of bits with the leading 1 bit as the sign bit $S$. By mapping the activations from the original \texttt{Linear} space to the \texttt{Log} space, the distribution of the activations is more uniform. To be specific, given a tile of elements, $[x_1,\cdots,x_m]$, which is 1x128 in our implementation, we take the absolute values and compute the logarithm of all the elements, and find the minimum $min=log(abs(x_i))$ and maximum $max=log(abs(x_j))$. The minimum is encoded as $S.00\cdots01$ and the maximum is encoded as $S.11\cdots11$, with an interval representing $Step=\frac{max-min}{2^{n-1}-2}$. Zero values are represented by $S.00\cdots00$, specially. The left values are rounded to the nearest integer $K$ multiples of $Step$. The decoding process is simple by combining the sign bit and $exp^{min+Step\times(K-1)}$. 

By locally calculating the $min$ and $Step$, this data type supports dynamic representation range for different blocks, covering larger ranges or providing more precision, compared to static floating point formats. 
Besides, we find it is important to round in the original \texttt{Linear} space, instead of the \texttt{Log} space, for the unbiased activation quantization. We also constrain the $min$ to be larger than $max-log(2^{32})$, which means that the max representation range is similar to E5, a floating point with 5 exponents. 
We validate our LogFMT-nBit on dense language models with around 7 billion parameters, by quantifying the output of the residual branch to simulate the \texttt{combine} stage in MoE models. When setting $n=8$, sharing the same bits with FP8, the LogFMT-8Bit shows superior training accuracy compared to E4M3 or E5M2. After increasing the $n$ to 10 bits, we find it's similar to the BF16 \texttt{combine} stage. 

\subsubsection{Limitations:}
The initial purpose of using LogFMT is to apply it to activations during transmission or near activation functions, as it offers higher precision than  FP8 with the same bit width. However, subsequent computations require reconversion to BF16 or FP8 to accommodate the Hopper GPU tensor cores' data type. Due to insufficient GPU bandwidth for log/exp operations and excessive register pressure during encode/decode, if encode/decode operations are fused with all-to-all communication, the overhead can be substantial (50\%$\sim$100\%). Therefore, although experimental results validate the effectiveness of this format, we do not employ it eventually. 

\vspace{-0.4em}
\subsubsection{Suggestions:}
Providing native support for compression and decompression units tailored to FP8 or custom precision formats represents a viable approach for future hardware. This could help minimize bandwidth requirements and streamline communication pipelines. The reduced communication overhead is particularly helpful in bandwidth-intensive tasks like MoE training.

\vspace{-0.4em}
\section{Interconnection Driven Design}
\label{sec:scale_up_optimizations}

\subsection{Current Hardware Architecture}

\begin{figure}[!t]
    \centering
    \includegraphics[width=1\linewidth]{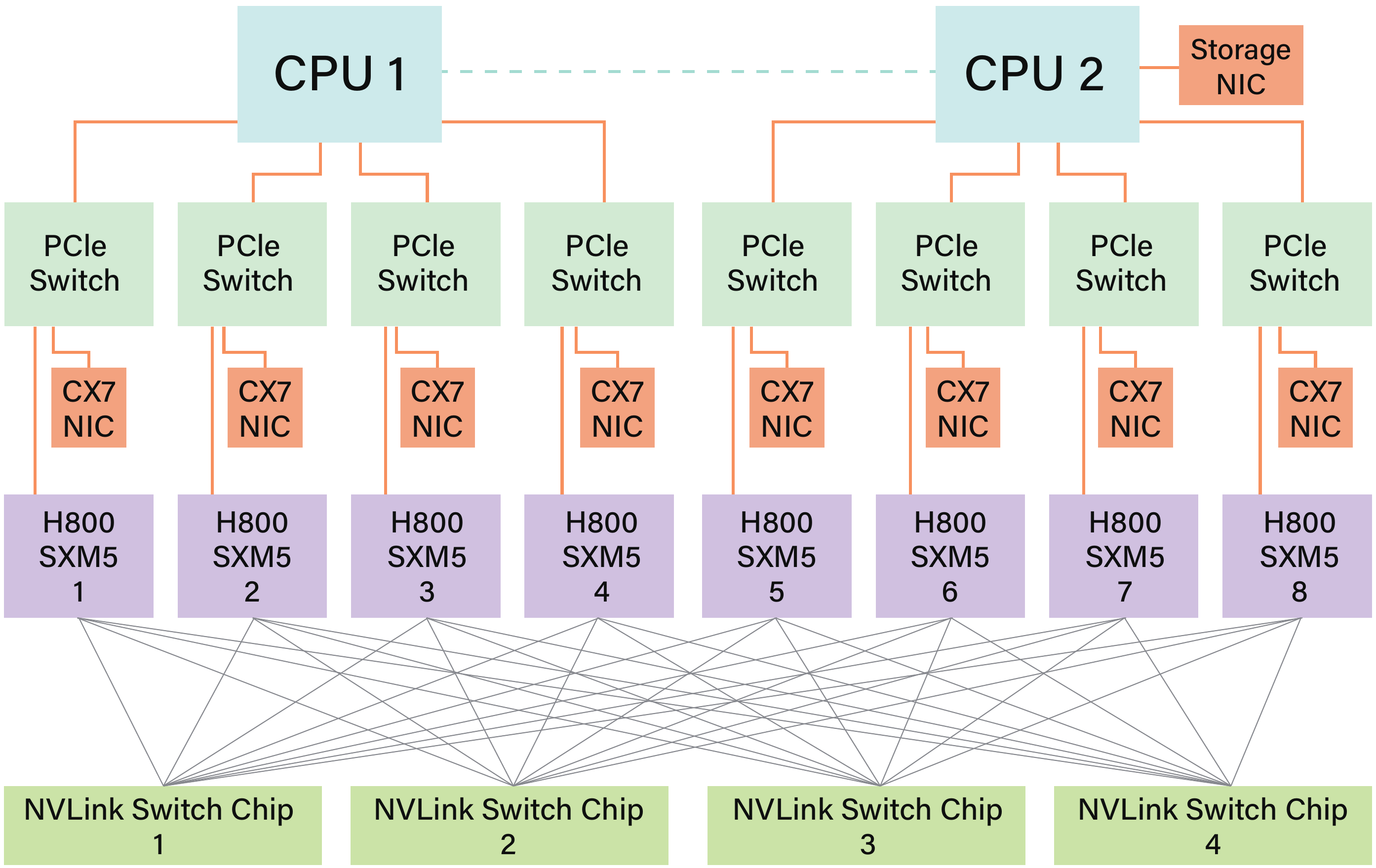}
    \vspace{-10pt}
    \caption{H800 node interconnection.}
    \label{fig:h800_arch}
    \vspace{-20pt}
\end{figure}

The NVIDIA H800 GPU SXM architecture we currently use, illustrated in Figure~\ref{fig:h800_arch}, is built on the Hopper architecture, similar to the H100 GPU. However, it features reduced FP64 computational performance and NVLink bandwidth for regulatory compliance. Specifically, the NVLink bandwidth in H800 SXM nodes is reduced from 900 GB/s to 400 GB/s. This significant reduction in intra-node scale-up bandwidth presents a challenge for high-performance workloads. To compensate, each node is equipped with eight 400G Infiniband (IB) CX7 NICs, enhancing scale-out capabilities to mitigate the bandwidth deficit.

To address these hardware constraints, the DeepSeek-V3 model incorporates several design considerations that align with the hardware's strengths and limitations.

\vspace{-0.5em}
\subsection{Hardware-Aware Parallelism}

To align with the constraints of the H800 architecture, the following parallelism strategies were considered to optimize the performance of DeepSeek-V3:

\begin{itemize}[leftmargin=10pt,topsep=0pt]
    \item \textbf{Avoidance of Tensor Parallelism (TP):} Tensor Parallelism is avoided during training due to its inefficiency under limited NVLink bandwidth. However, during inference, TP can still be selectively used to improve TTFT and TPOT performance.
    \item \textbf{Enhanced Pipeline Parallelism (PP):} DualPipe~\citep{dualpipe} is employed to overlap attention and MoE computation with MoE communication. This also reduces pipeline bubbles and balances memory usage across GPUs, improving overall throughput. Additional details are available in the technical report~\citep{dsviii}.
    \item \textbf{Accelerated Expert Parallelism (EP):} With eight 400Gbps InfiniBand (IB) NICs, the system achieves all-to-all communication at speeds exceeding 40GB/s. Notably, our all-to-all EP implementation, DeepEP~\citep{deepep2025}, is open-sourced, enabling highly efficient expert parallelism as discussed in the following subsection.
\end{itemize}

\subsection{Model Co-Design: Node-Limited Routing}
The bandwidth disparity between scale-up (intra-node) and scale-out (inter-node) communication in the H800 architecture is approximately 4:1. Specifically, NVLink provides 200GB/s  bandwidth (of which about 160GB/s can actually be achieved), while each 400Gbps IB NIC delivers only 50GB/s bandwidth (we consider small message size and latency influence, use 40GB/s for effective bandwidth). To balance and fully utilize the higher intra-node bandwidth, the model architecture is co-designed with hardware, particularly in the \textbf{TopK Expert Selection Strategy}.

Consider a setup with 8 nodes (64 GPUs in total) and 256 routed experts (4 experts per GPU). 
For \dsviii{}, each token is routed to one shared expert and 8 routed experts. 
If its 8 target experts are distributed across all 8 nodes, the communication time over IB would be $8t$, where $t$ represents the time to send one token over IB. 
However, by leveraging the higher NVLink bandwidth, tokens routed to the same node can be sent once over IB and then forwarded via NVLink to other intra-node GPUs. 
The NVLink forwarding enables deduplication of the IB traffic. 
When the target experts for a given token are distributed across $M$ nodes, the deduplicated IB communication cost will be reduced to $Mt$ ($M < 8$).

Since the IB traffic depends on only $M$, \dsviii{} introduces a \textbf{Node-Limited Routing} for the TopK expert selection strategy. 
Specifically, we group 256 routed experts into 8 groups, with 32 experts per group, and deploy each group on a single node.
On top of this deployment, we algorithmically ensure that each token will be routed to up to 4 nodes. This approach mitigates the bottleneck of IB communication and enhances the effective communication bandwidth during training.

\vspace{-0.3em}
\subsection{Scale-Up and Scale-Out Convergence}

\subsubsection{Limitations of Current Implementations}

While the Node-Limited Routing strategy reduces communication bandwidth requirements, it complicates communication pipeline kernel implementations due to the disparity in bandwidth between intra-node (NVLink) and inter-node (IB) interconnects. 
In practice, GPU Streaming Multiprocessors (SM) threads are used for both network message handling (e.g., filling QPs and WQEs) and data forwarding over NVLink, consuming computational resources. For example, during training, up to 20 of the SMs on the H800 GPU are allocated for communication-related operations, leaving fewer resources available for actual computation.
To maximize throughput in online inference, we perform EP all-to-all communication entirely through NIC RDMA, avoiding SM resource contention and improving compute efficiency. This highlights the advantage of RDMA’s asynchronous communication model in overlapping computation and communication.

The following are key tasks currently performed by SMs during EP communication, particularly for the \texttt{combine} stage’s \texttt{reduce} operations and data type conversions. Offloading these tasks to dedicated communication hardware could free up SMs for computation kernels, significantly improving overall efficiency:
\begin{itemize}[leftmargin=10pt,topsep=0pt]
    \item \textbf{Forwarding Data:} Aggregating IB traffic destined for multiple GPUs within the same node between the IB and NVLink domains.
    \item \textbf{Data Transport:} Moving data between RDMA buffers (registered GPU memory regions) and input/output buffers.
    \item \textbf{Reduce Operations:} Executing \texttt{reduce} operations required for EP \texttt{all-to-all} \texttt{combine} communications.
    \item \textbf{Managing Memory Layouts:} Handling fine-grained memory layouts for chunked data transfers across the IB and NVLink domains.
    \item \textbf{Data Type Cast}: Converting data type before and after \texttt{all-to-} \texttt{all} communications.
\end{itemize}
\vspace{-0.8em}

\subsubsection{Suggestions:}
To address these inefficiencies, we strongly recommend that future hardware should integrate intra-node (scale-up) and inter-node (scale-out) communication into a unified framework. By incorporating dedicated co-processors for network traffic management and seamless forwarding between NVLink and IB domains, such designs can reduce software complexity and maximize bandwidth utilization. For example, node-limited routing strategies employed in DeepSeek-V3 can be further optimized with hardware support for dynamic traffic deduplication.

We also recognize emerging interconnect protocols such as the Ultra Ethernet Consortium (UEC)~\citep{uec_overview,uec_update1}, Ultra Accelerator Link (UALink)~\citep{ualink_white_paper}, both of which are poised to drive advancements in scale-up and scale-out communication. More recently, Unified Bus (UB)~\citep{liao2025ubmeshhierarchicallylocalizedndfullmesh} has introduced a novel approach to scale-up and scale-out convergence. Section~\ref{sec:future_hardware} further explores several technical innovations proposed by UEC and UALink. However, in this section, our primary focus is on achieving scale-up and scale-out convergence at the programming framework level:

\begin{enumerate}[leftmargin=12pt,topsep=0pt]
    \item \textbf{Unified Network Adapter:} Design NICs (Network Interface Cards) or I/O Dies that are connected to unified scale-up and scale-out networks. These adapters should also support basic switch functionality, such as forwarding packets from the scale-out network to specific GPUs within the scale-up network. This could be achieved using a single LID (Local Identifier) or IP address with policy-based routing.
    \item \textbf{Dedicated Communication Co-Processor:} Introduce a dedicated co-processor or programmable component—such as an I/O die—for handling network traffic. This component should offload packet processing from GPU SMs, provide hardware-accelerated memory copy for efficient buffer management, and, crucially, accelerate memory load/store operations in a manner similar to TMA (Tensor Memory Accelerator), thereby saturating bandwidth with minimal resource consumption.
    \item \textbf{Flexible Forwarding, Broadcast and Reduce Mechanisms:} Hardware should support flexible forwarding, broadcast operations (for EP dispatch), and reduce operations (for EP combine) across scale-up and scale-out networks—mirroring our current GPU SM-based implementation. This would not only improve effective bandwidth but also reduce the computational complexity of network-specific operations. 
    \item \textbf{Hardware Synchronization Primitives:} Provide fine-grained hardware synchronization instructions to handle memory consistency issues or out-of-order packet arrivals at the hardware level. This would eliminate the need for software-based synchronization mechanisms like RDMA completion events, which introduce extra latency and increase programming complexity. Memory-semantic communication with an acquire/release mechanism is a promising implementation.
\end{enumerate}

By implementing these recommendations, future hardware designs can significantly enhance the efficiency of large-scale distributed AI systems while simplifying software development.


\subsection{Bandwidth Contention and Latency}
\label{section:innode_bw_contention}

\subsubsection{Limitations:}
Besides, current hardware lacks the flexibility to dynamically allocate bandwidth between different types of traffic on NVLink and PCIe. For example, during inference, transferring KV cache data from CPU memory to GPU can consume tens of GB/s, saturating PCIe bandwidth. If the GPU simultaneously uses IB for EP communication, this contention between KV cache transfers and EP communication can degrade overall performance and cause latency spikes.

\begin{figure}[!b]
    \centering
    \includegraphics[width=\linewidth]{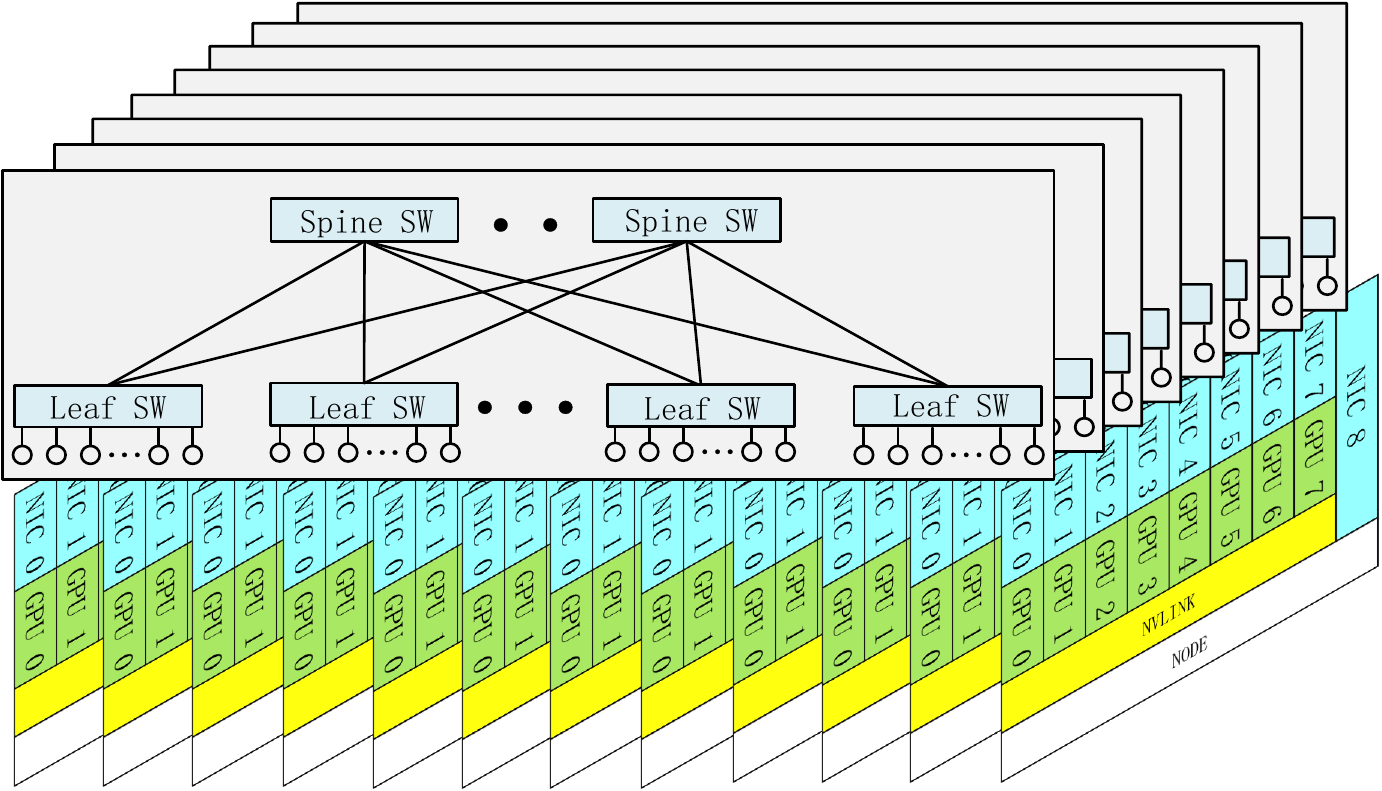}
    \vspace{-15pt} 
    \caption{Eight-plane two-layer fat-tree scale-out network: Each GPU and IB NIC pair belongs to one network plane. Cross-plane traffic must use another NIC and PCIe or NVLink for intra-node forwarding.}
    \label{nextgen}
\end{figure}

\begin{figure*}[!t]
    \centering
    \includegraphics[width=\linewidth]{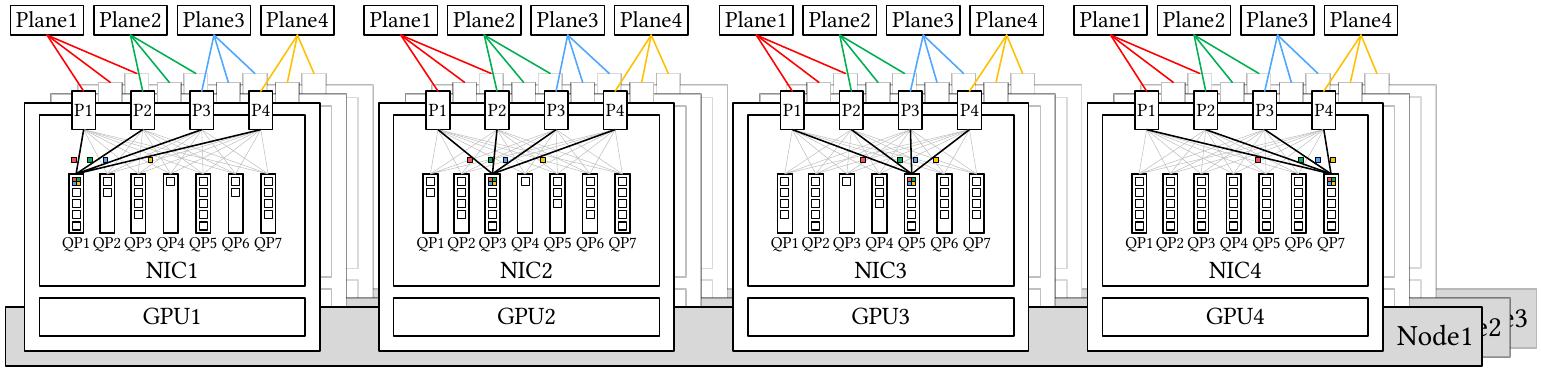}
    \vspace{-5pt} 
    \caption{Ideal Multi-Plane Network: Each NIC is equipped with multiple physical ports, each connected to a distinct network plane. A single queue pair (QP) can simultaneously utilize all available ports for transmitting and receiving packets, which necessitates native support for out-of-order placement within the NIC. 
    }
    \label{nextgen_mp}
    \vspace{-5pt} 
\end{figure*}

\vspace{-0.4em}
\subsubsection{Suggestions:}
\begin{itemize}[leftmargin=10pt,topsep=0pt]
    \item \textbf{Dynamic NVLink/PCIe Traffic Prioritization:} Hardware should support dynamic prioritization of traffic based on its type. For example, traffic related to EP, TP, and KV cache transfers should be assigned different priorities to maximize interconnect efficiency. For PCIe, exposing the traffic class (TC) to user-level programming would suffice.
    \item \textbf{I/O Die Chiplet Integration:} Integrating NICs directly into the I/O die and connecting them to the compute die in the same package, rather than through conventional PCIe, would substantially reduce communication latency and alleviate PCIe bandwidth contention. 
    \item \textbf{CPU–GPU Interconnects within the Scale-Up Domain:} To further optimize intra-node communication, CPUs and GPUs should be interconnected using NVLink or similar dedicated high-bandwidth fabrics, rather than relying solely on PCIe. Similar to the benefits provided by integrating NICs into the I/O die, this approach can significantly improve scenarios such as offloading parameters or KV cache between GPU and CPU memory during training and inference.
\end{itemize}

\vspace{-1em}

 \vspace{-0.5em}
\section{Large Scale Network Driven Design}
\label{sec:scale_out_optimizations}
\subsection{Network Co-Design: Multi-Plane Fat-Tree}

During the training of DeepSeek-V3, we deployed a \textbf{Multi-Plane Fat-Tree (MPFT)} scale-out network, as shown in Figure~\ref{nextgen}. Each node is equipped with eight GPUs and eight IB NICs, with each GPU–NIC pair assigned to a distinct network plane. Additionally, each node has a 400\,Gbps Ethernet RoCE NIC connected to a separate storage network plane for accessing the 3FS~\citep{ds3fs} distributed file system. In the scale-out network, we used 64-port 400G IB switches, enabling the topology theoretically supports up to 16,384 GPUs while retaining the cost and latency advantages of a two-layer network. However, due to policy and regulatory constraints, just over two thousand GPUs were ultimately deployed.

Furthermore, due to the current limitations of IB ConnectX-7, our deployed MPFT network does not fully realize the envisioned architecture. Ideally, as depicted in Figure~\ref{nextgen_mp}, each NIC would feature multiple physical ports, each connected to a separate network plane, yet collectively exposed as a single logical interface to the user through port bonding. From a user perspective, a single Queue Pair (QP) could seamlessly transmit and receive messages across all available ports, akin to packet spraying. As a consequence, packets originating from the same QP may traverse distinct network paths and arrive at the receiver out of order, thereby necessitating native support for out-of-order placement within the NIC to guarantee message consistency and preserve the correct ordering semantics. For example, InfiniBand ConnectX-8 natively supports four plane. It would be advantageous for future NICs to fully support advanced multi-plane capabilities, allowing two-tier fat-tree networks to scale effectively to much larger AI clusters. Overall, the multi-plane architecture offers significant advantages in fault isolation, robustness, load balancing, and large-scale system scalability.

\subsubsection{Advantages of Multi-Plane Fat-Tree Network}
\begin{itemize}[leftmargin=10pt,topsep=0pt]
    \item \textbf{Subset of Multi-Rail Fat-Tree (MRFT):} The MPFT topology constitutes a specific subset of the broader MRFT architecture. As a result, existing optimizations developed by NVIDIA and NCCL for Multi-Rail networks can be seamlessly leveraged within Multi-Plane network deployments. Furthermore, NCCL’s support for PXN~\cite{nccl_pxn} technology addresses the inherent challenge of inter-plane isolation, enabling efficient communication even when direct interconnectivity between planes is absent.
    \item \textbf{Cost Efficiency:} As shown in Table~\ref{tab:network_topology_comparison}, the multi-plane network enables over 10k endpoints using a two-layer fat-tree (FT2) topology, significantly reducing network costs compared to a three-layer fat tree (FT3).  The cost per endpoint is even slightly more competitive than the cost-efficient Slim Fly (SF) topology~\cite{10.5555/3691825.3691882}.
    \item \textbf{Traffic Isolation:} Each plane operates independently, ensuring that congestion in one plane does not affect others. This isolation improves overall network stability and prevents cascading performance degradation.
    \item \textbf{Latency Reduction:} The two-layer topology achieves lower latency than three-layer fat trees, as demonstrated in our experiments. This makes it particularly suitable for latency-sensitive applications such as MoE-based training and inference.
    \item \textbf{Robustness:} As shown in Figure~\ref{nextgen_mp}, multi-port NICs provide multiple uplinks, so single-port failures do not disrupt connectivity and rapid, transparent fault recovery is possible.

\end{itemize}

\begin{table}[!t]
\centering
\small
\caption{Network topology comparison. Cost estimates are derived from the methodology in the Slim Fly (SF) paper~\cite{10.5555/3691825.3691882}. DF denotes the canonical dragonfly topology~\citep{9355230,4556717,10.1145/3295500.3356208}.}
\vspace{-5pt}
\begin{tabular}{|l|c|c|c|c|c|}
\hline
\textbf{Metric}         & \textbf{FT2} & \textbf{MPFT} & \textbf{FT3} & \textbf{SF} & \textbf{DF} \\ \hline
Endpoints               & 2,048        & 16,384        & 65,536       & 32,928      & 261,632     \\ \hline
Switches                & 96           & 768           & 5,120        & 1,568       & 16,352      \\ \hline
Links                   & 2,048        & 16,384        & 131,072      & 32,928      & 384,272     \\ \hline
Cost [M\$]              & 9            & 72            & 491          & 146         & 1,522       \\ \hline
Cost/Endpoint [k\$]     & 4.39         & 4.39          & 7.5          & 4.4         & 5.8         \\ \hline
\end{tabular}
\label{tab:network_topology_comparison}
\end{table}

It is important to note that, due to current 400G NDR InfiniBand limitations, cross-plane communication requires intra-node forwarding, which introduces additional latency during inference. If future hardware can achieve scale-up and scale-out network convergence as discussed earlier, this latency can be significantly reduced, further enhancing the viability of multi-plane networks.

\begin{figure}[!t]
\centering
\centerline{}\includegraphics[width=1\linewidth]{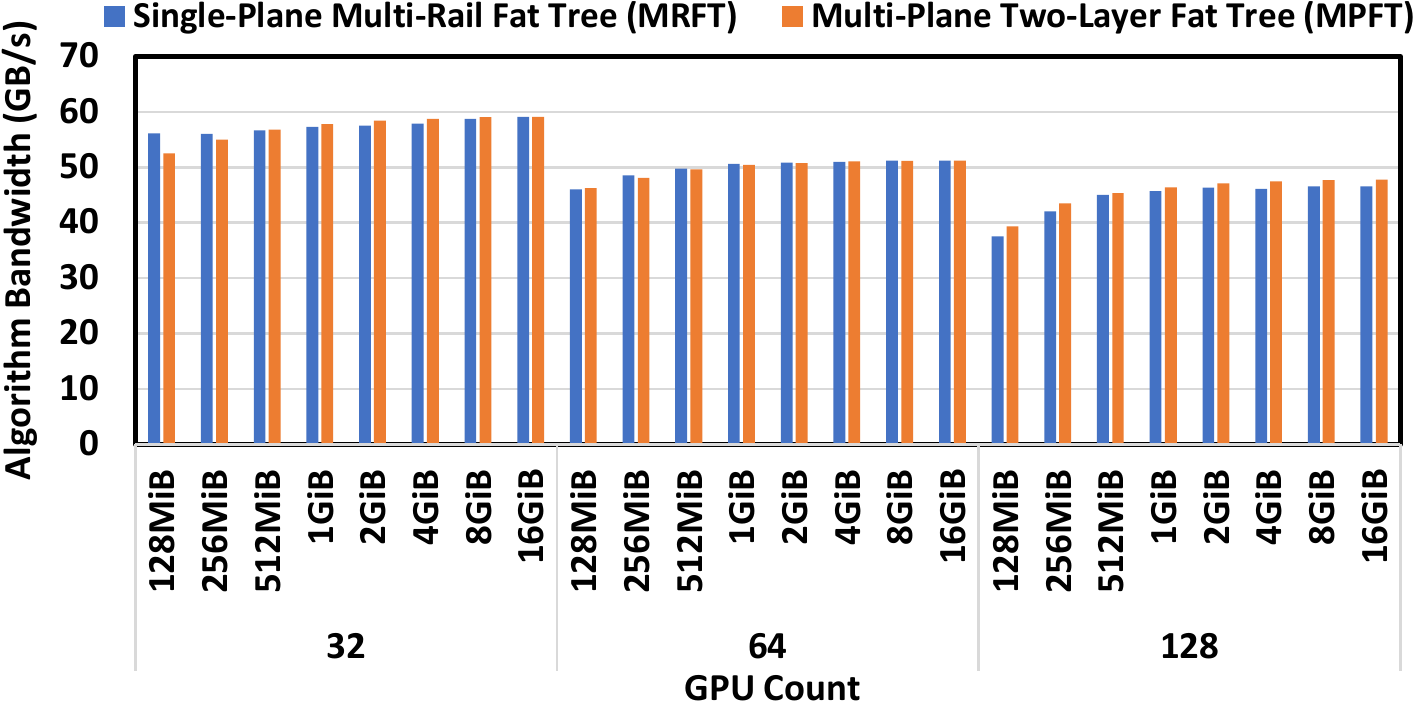}
\vspace{-15pt}
\caption{NCCL all-to-all performance from 32 to 128 GPUs for MRFT and MPFT networks.}
\vspace{-5pt}
\label{fig:nccl_allgather}
\end{figure}

\begin{figure}[bt]
    \centering
    \includegraphics[width=\linewidth]{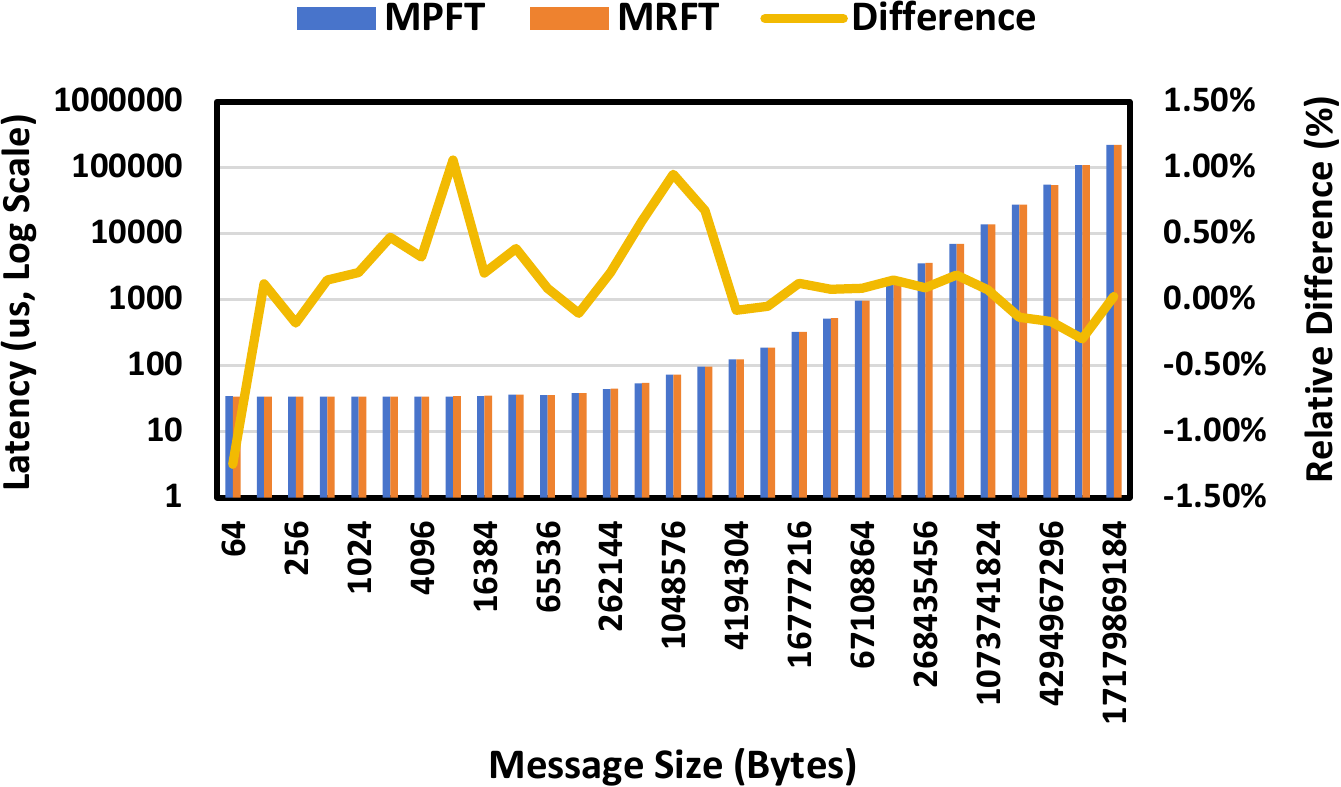}
    \vspace{-5pt} 
    \caption{Latency comparison between MPFT and MRFT networks in NCCL all-to-all test under different message sizes, showing that their performance is nearly identical.}
    \label{fig:alltoall_latency_comparison}
\end{figure}

\subsubsection{Performance Analysis}

To verify the effectiveness of the Multi-Plane Network design, we conducted real-world experiments on our cluster, modifying the cluster's network topology to compare the performance of the \textbf{Multi-Plane Two-Layer Fat Tree (MPFT)} and the \textbf{Single-Plane Multi-Rail Fat Tree (MRFT)}. Below are the key findings from our experiments:

1. \textbf{All-to-All Communication and EP Scenarios}:  
As illustrated in Figure~\ref{fig:nccl_allgather}, the all-to-all performance of the multi-plane network is very similar to that of the single-plane multi-rail network. This performance parity can be attributed to NCCL's PXN \cite{nccl_pxn} mechanism, which optimizes traffic forwarding via NVLink in multi-rail topologies. The multi-plane topology also benefits from this mechanism. As shown in Figure~\ref{fig:alltoall_latency_comparison}, the results of all-to-all communication tests conducted on 16 GPUs reveal negligible differences in latency between the MPFT and MRFT topologies.

\begin{figure}[bt]
    \centering
    \includegraphics[width=\linewidth]{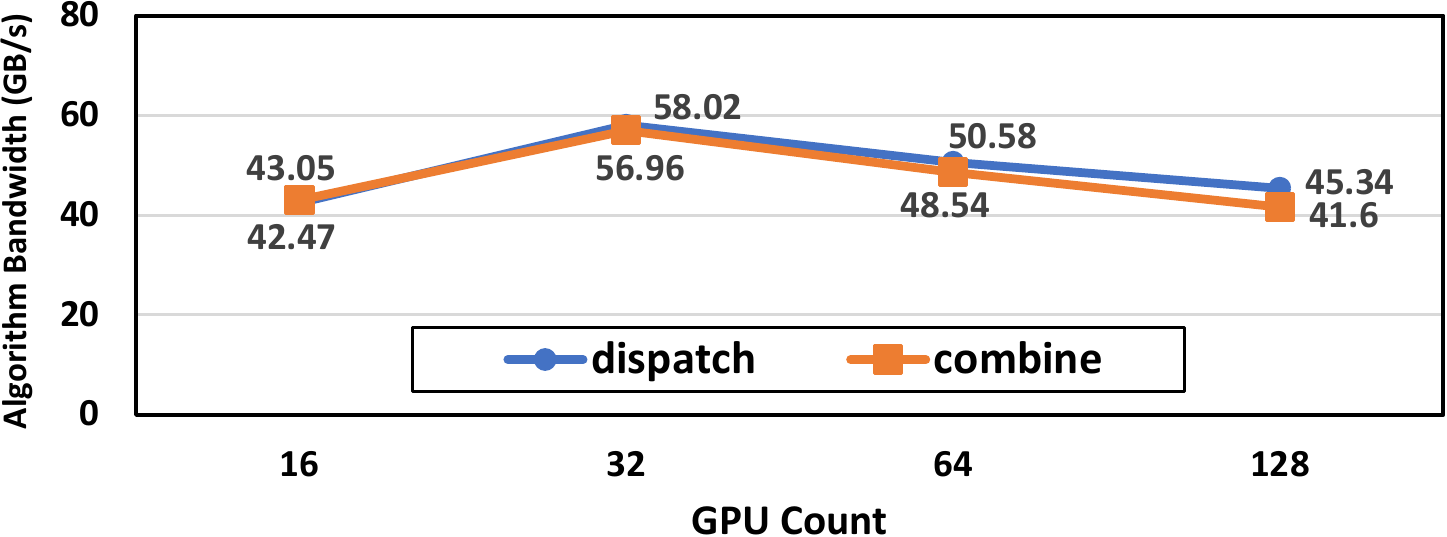}
    \vspace{-15pt} 
    \caption{DeepEP performance on MPFT: The EP dispatch and combine kernel communicates across 16 to 128 GPUs using all-to-all. Each GPU processes 4096 tokens. The observed throughput nearly saturates the 400Gps NIC bandwidth.}
    \label{alltoall_comparison}
    \vspace{-5pt} 
\end{figure}

To evaluate MPFT's performance of all-to-all communication in practical training scenarios, we tested the EP communication patterns commonly used during training. As shown in Figure~\ref{alltoall_comparison}, each GPU achieves a high bandwidth exceeding 40GB/s in a multi-plane network, providing reliable performance that meets the demands of training.

2. \textbf{Training Throughput for DeepSeek-V3 Model}:  
   We also compare the training metrics of the DeepSeek-V3 model between MPFT and MRFT in Table~\ref{tab:training_comparison}.
   MFU (Model Flops Utilization) is calculated based on BF16 peak performance. Causal MFU only takes into account the flops of the lower triangle of the attention matrix (in line with FlashAttention\citep{dao2022flashattention, dao2023flashattention2}), while non-causal MFU includes the flops of the whole attention matrix (in line with Megatron~\citep{megatron3}). 1F, 1B, and 1W denote forward time, input backward time, and weight backward time, respectively. 
   When training the V3 model on 2048 GPUs, the performance of MPFT is nearly identical to that of MRFT, with observed differences falling within normal fluctuations and measurement error.


\subsection{Low Latency Networks}

In our model inference, large-scale EP relies heavily on \texttt{all-to-all} communication, which is highly sensitive to both bandwidth and latency. Consider a typical scenario discussed in Section~\ref{section:inference_speed_limits}, with a network bandwidth of 50GB/s, the data transfer should ideally take approximately $120~\mu\mathrm{s}$
. Therefore, the intrinsic network latencies on the order of microseconds can critically impact system performance, making their effects non-negligible.

\subsubsection{IB or RoCE}

\begin{table}[!t]
\centering
\small
\caption{Training metric comparison between MPFT and MRFT networks.}
\label{tab:training_comparison}
\vspace{-5pt}
\begin{tabular}{|l|c|c|}
\hline
\textbf{Metric}                     & \textbf{MPFT} & \textbf{MRFT} \\ \hline
\textbf{tokens/day (B)}             & 272.80            & 272.52             \\ \hline
\textbf{time/step (s)}              & 19.926            & 19.946             \\ \hline
\textbf{1F (s)}                     & 1.13          & 1.13          \\ \hline
\textbf{bubble (s)}                 & 2.06          & 2.03          \\ \hline
\textbf{1B (s)}                     & 1.99          & 1.99          \\ \hline
\textbf{1W (s)}                     & 0.48          & 0.48          \\ \hline
\textbf{1F1B (s)}                   & 13.95         & 14.00         \\ \hline
\textbf{opt (s)}                    & 0.29          & 0.31          \\ \hline
\textbf{TFLOPS (non-causal)}        & 432               & 432                \\ \hline
\textbf{TFLOPS (causal)}            & 385               & 385                \\ \hline
\textbf{MFU (non-causal)}      & 43.73\%             & 43.68\%              \\ \hline
\textbf{MFU (causal)}          & 38.94\%             & 38.90\%              \\ \hline
\end{tabular}
\vspace{-5pt}
\end{table}

As shown in Table \ref{tab:latency_comparison}, IB consistently achieves lower latency, making it the preferred choice for latency-sensitive workloads such as distributed training and inference. Although IB has superior latency performance compared to RDMA over Converged Ethernet (RoCE), it comes with certain limitations:
\begin{itemize}
    \item \textbf{Cost:} IB hardware is significantly more expensive than RoCE solutions, which limits its widespread adoption.
    \item \textbf{Scalability:} IB switches typically support only 64 ports per switch, compared to the 128 ports commonly found in RoCE switches. This restricts the scalability of IB-based clusters, particularly for large-scale deployments.
\end{itemize}

\begin{table}[!t]
    \centering
    \small
    \caption{CPU side end-to-end latency comparison between IB, RoCE, and intra-node NVLink for 64B data transmission.}
    \label{tab:latency_comparison}
    \vspace{-5pt}
    \begin{tabular}{|c|c|c|c|}
        \hline
        \textbf{Link Layer} & \textbf{Same Leaf} & \textbf{Cross Leaf} \\
        \hline
        RoCE  & 3.6us & 5.6us \\
        \hline
        InfiniBand & 2.8us & 3.7us \\
        \hline
        NVLink & 3.33us & - \\
        \hline
    \end{tabular}
    \vspace{-5pt} 
\end{table}

\subsubsection{Recommendations for RoCE Improvements}
\label{section:future_network_recommendations}
\begin{figure}[!t]
    \centering
    \includegraphics[width=0.475\textwidth]{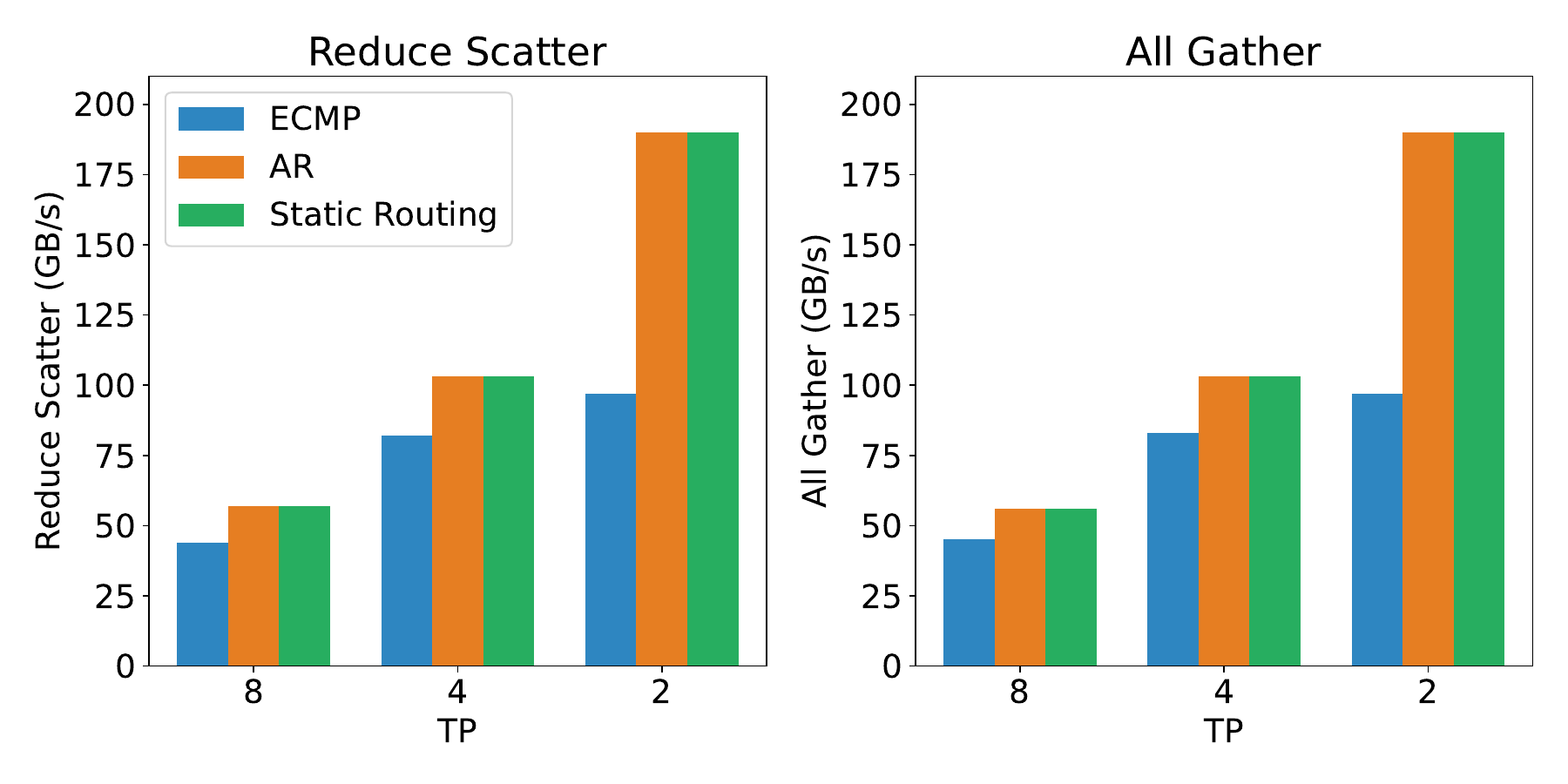} 
    \vspace{-10pt} 
    \caption{RoCE network bandwidth of AllGather and ReduceScatter communication primitives under different routing methods (ECMP, AR, Static Routing) and TP dimensions.}
    \label{fig:route_perf}
       \vspace{-10pt} 
\end{figure}

While RoCE has the potential to be a cost-effective alternative to IB, its current limitations in latency and scalability prevent it from fully meeting the demands of large-scale AI systems. Below, we outline specific recommendations for improving RoCE:

\begin{enumerate}[leftmargin=12pt,topsep=0pt]
    \item \textbf{Specialized Low-Latency RoCE Switches:} 
    We recommend that Ethernet vendors develop RoCE switches specifically optimized for RDMA workloads by removing unnecessary Ethernet features. The Slingshot architecture~\citep{9355230} exemplifies how Ethernet-based designs can achieve latency performance comparable to IB. Similarly, recent innovations from Broadcom~\citep{brcm_eth_scale}, including the AI Forwarding Header (AIFH) and upcoming low-latency Ethernet switches, demonstrate the feasibility of high-performance Ethernet fabrics tailored for AI. We are looking forward to continuing innovation in this direction.

   \item \textbf{Optimized Route Policy:}  As shown in Figure \ref{fig:route_perf}, the default Equal-Cost Multi-Path (ECMP) routing policy in RoCE struggles to distribute traffic efficiently across interconnects, leading to severe congestion performance degradation in NCCL collective communication tests.  LLM training traffic, such as in DP (Data Parallelism), tends to lack randomness, causing multiple flows to converge on the same interconnect link. In contrast, Adaptive Routing (AR)~\citep{4618589} can significantly enhance network performance by dynamically spraying packets across multiple paths. While static routing—based on manually configured route tables—can avoid link conflicts for specific destinations, it lacks flexibility. For large-scale all-to-all communication, adaptive routing offers superior performance and scalability.

    \item \textbf{Improved Traffic Isolation or Congestion Control Mechanisms:} 
    Current RoCE switches support only a limited number of priority queues, which are insufficient for complex AI workloads involving concurrent communication patterns such as EP’s all-to-all and DP’s all-reduce. In such mixed workloads, all-to-all traffic can cause incast congestion due to bursty many-to-one transfers, potentially degrading overall network performance. To address incast's influence on other traffic, one approach is to adopt virtual output queuing (VOQ), assigning a dedicated virtual queue to each QP to isolate traffic flows. Alternatively, more effective congestion control (CC) mechanisms such as RTT-based CC (RTTCC) or user-programmable CC (PCC) can be employed, enabling NIC–switch co-optimization to maintain low latency and high throughput under dynamic traffic conditions.

\end{enumerate}

\subsubsection{InfiniBand GPUDirect Async (IBGDA)}

We utilize IBGDA~\citep{nvshmem_ibgda,AGOSTINI201828} to reduce latency in network communications. Traditionally, network communication involves the creation of a CPU proxy thread: once the GPU has prepared the data, it must notify the CPU proxy, which then populates the control information for the work request (WR) and signals the NIC via a doorbell mechanism to initiate data transmission. This process introduces additional communication overhead.

IBGDA addresses this issue by allowing the GPU to directly fill the WR content and write to the RDMA doorbell MMIO address. By managing the entire control plane within the GPU, IBGDA eliminates the significant latency overhead associated with GPU-CPU communication. Moreover, when sending a large number of small packets, the control plane processor can easily become a bottleneck. Since GPUs have multiple parallel threads, the sender can leverage these threads to distribute the workload, thereby avoiding such bottlenecks. A range of works—including our DeepEP~\citep{deepep2025}—have leveraged IBGDA and reported substantial performance gains~\citep{7973709,chen2025efficientheterogeneouslargelanguage,zheng2025tilelinkgeneratingefficientcomputecommunication}. We therefore advocate for such capabilities to be widely supported across accelerator devices.

\section{Discussion and Insights for Future Hardware Architecture Design}
\label{sec:future_hardware}
Building on the previous sections, we summarize key architectural insights and outline future directions for hardware design tailored to large-scale AI workloads. 

Section~\ref{section:inference_speed_limits} highlighted the importance of large-scale scale-up networks for accelerating model inference. Section~\ref{sec:hardware_features} discussed the necessity of efficient support for low-precision computation and communication. Section~\ref{sec:scale_up_optimizations} explored the convergence of scale-up and scale-out architectures, along with several proposed enhancements. Section~\ref{sec:scale_out_optimizations} focused on multi-plane network topologies and identified key improvements needed for Ethernet-based interconnects.

Together, these sections identify hardware limitations in concrete application contexts and offer corresponding suggestions. Building on that foundation, this section expands the discussion to broader considerations and proposes forward-looking directions for future hardware architecture design.

\subsection{Robustness Challenges}

\subsubsection{Limitations:}

\begin{itemize}[leftmargin=10pt,topsep=0pt]
    \item \textbf{Interconnect Failures:} High-performance interconnects (e.g., IB and NVLink) are prone to intermittent disconnections, which can disrupt node-to-node communication. This is especially harmful in communication-heavy workloads like EP, where even brief interruptions may lead to significant performance drops or job failures.

    \item \textbf{Single Hardware Failures:} Node crashes, GPU failures, or ECC (Error-Correcting Code) memory errors can compromise long-running training jobs, often requiring costly restarts. The impact of such failures escalates in large-scale deployments, where the probability of a single-point failure increases proportionally with system size.

    \item \textbf{Silent Data Corruption:} Errors undetected by ECC mechanisms, such as multi-bit memory flips or computational inaccuracies, pose a significant risk to model quality. These errors are particularly insidious in long-running tasks, as they can propagate undetected and corrupt downstream computations. Current mitigation strategies rely on application-level heuristics, which are insufficient for ensuring system-wide robustness.
\end{itemize}

\subsubsection{Suggestions for Advanced Error Detection and Correction}
To mitigate risks associated with silent corruption, hardware must incorporate advanced error detection mechanisms beyond traditional ECC. Techniques such as checksum-based validation or hardware-accelerated redundancy checks can provide higher reliability for large-scale deployments.

Furthermore, hardware vendors should deliver comprehensive diagnostic toolkits to end users, empowering them to rigorously verify the integrity of their systems and proactively identify any latent silent data corruption. Such toolkits, when embedded as part of the standard hardware package, foster transparency and enable continuous validation throughout the operational lifecycle, thereby bolstering overall system trustworthiness.

\subsection{CPU Bottlenecks and Interconnects}

While accelerator design often takes center stage, CPUs remain essential for coordinating computation, managing I/O, and sustaining system throughput. However, current architectures face several critical bottlenecks:

First, as discussed in Section~\ref{section:innode_bw_contention}, the PCIe interface between CPUs and GPUs often becomes a bandwidth bottleneck, particularly during large-scale parameter, gradient, or KV cache transfers. To mitigate this, future systems should adopt direct CPU–GPU interconnects—such as NVLink or Infinity Fabric—or integrate both CPUs and GPUs into the scale-up domain, thereby eliminating intra-node bottlenecks.

In addition to PCIe limitations, sustaining such high data transfer rates also requires exceptionally high memory bandwidth. For example, saturating 160 lanes of PCIe 5.0 demands over 640~GB/s per node, translating to a memory bandwidth requirement of approximately 1~TB/s per node—posing a significant challenge for conventional DRAM architectures.

Lastly, latency-sensitive tasks such as kernel launches and network processing demand high single-core CPU performance, typically requiring base frequencies above 4~GHz. Furthermore, modern AI workloads require sufficient CPU cores per GPU to prevent control-side bottlenecks. For chiplet-based architectures, additional cores are needed to support cache-aware workload partitioning and isolation.

\subsection{Toward Intelligent Networks for AI}
To meet the demands of latency-sensitive workloads, future interconnects must prioritize both low latency and intelligent networks:
\begin{itemize}[leftmargin=10pt,topsep=0pt]
    \item \textbf{Co-Packaged Optics:} Incorporating silicon photonics enables scalable higher bandwidth scalability and enhanced energy efficiency, both are critical for large-scale distributed systems.
    \item \textbf{Lossless Network}: Credit-Based Flow Control (CBFC) mechanisms ensures lossless data transmission, yet naively triggering flow control can induce severe head-of-line blocking. Therefore, it is imperative to deploy advanced, endpoint-driven congestion control (CC) algorithms that proactively regulate injection rates and avert pathological congestion scenarios.
    \item \textbf{Adaptive Routing:}  As underscored in Section~\ref{section:future_network_recommendations}, future network should standardize the adoption of dynamic routing schemes—such as packet spraying and congestion-aware path selection—that continuously monitor real-time network conditions and intelligently redistribute traffic. These adaptive strategies are particularly effective in alleviating hotspots and mitigating bottlenecks during collective communication workloads, including all-to-all and reduce-scatter operations.
    \item \textbf{Efficient Fault-Tolerant Protocols:} Robustness against failures can be significantly enhanced through the deployment of self-healing protocols, redundant ports, and rapid failover techniques. For instance, link-layer retry mechanisms and selective retransmission protocols prove indispensable in scaling reliability across large networks, minimizing downtime and ensuring seamless operation despite intermittent failures.
    \item \textbf{Dynamic Resource Management:} To handle mixed workloads effectively, future hardware should enable dynamic bandwidth allocation and traffic prioritization. For example, inference tasks should be isolated from training traffic in unified clusters, ensuring responsiveness for latency-sensitive applications.
\end{itemize}


\subsection{Discussion on Memory-Semantic Communication and Ordering Issue }
Inter-node communication using load/store memory semantics is efficient and programmer-friendly, but current implementations are hampered by memory ordering challenges. For example, after writing data, the sender must issue an explicit memory barrier (fence) before updating a flag to notify the receiver, ensuring data consistency. This strict ordering introduces additional round-trip time (RTT) latency and can stall the issuing thread, impeding inflight stores and reducing throughput. Similar out-of-order synchronization issues arise in message-semantic RDMA; for instance, performing RDMA atomic add operations with packet spraying after regular RDMA writes on InfiniBand or NVIDIA BlueField-3 can incur additional RTT latency.

To address these, we advocate for hardware support that offers built-in ordering guarantees for memory-semantic communication. Such consistency should be enforced both at the programming level (e.g., via acquire/release semantics) and by hardware at the receiver, enabling in-order delivery without added overhead.

Several approaches are possible. For example, the receiver can buffer atomic messages and use packet sequence numbers (PSN) to ensure in-order processing; this method is straightforward and effective for maintaining correctness. Alternatively, a region-based acquire/release (RAR) mechanism is particularly attractive, as it enables true acquire/release semantics for remote memory access and offers greater flexibility for a range of workloads. In this approach, the receiver hardware maintains lightweight metadata—such as bitmaps or region-based counters—to track the state of memory regions. Acquire and release operations are scoped to specific address ranges, enabling efficient, hardware-enforced ordering without explicit sender-side fences. Notably, both approaches are amenable to implementation on the NIC or I/O die and are applicable to both memory-semantic and message-semantic RDMA primitives, thereby broadening their practical utility.


\subsection{In-Network Computation and Compression}
EP involves two critical \texttt{all-to-all} stages—\textbf{dispatch} and \textbf{combine}—that present significant opportunities for in-network optimization. The \textbf{dispatch} stage resembles a small-scale multicast operation, where a single message must be forwarded to multiple target devices. A hardware-level protocol enabling automatic packet replication and forwarding to multiple destinations could drastically reduce communication overhead and improve efficiency.

The \textbf{combine} stage, acting as a small-scale reduction operation, could benefit from in-network aggregation techniques. However, due to the small reduction scope and imbalanced workload in EP combine, implementing in-network aggregation in a flexible manner is challenging.

Moreover, as highlighted in Section \ref{sec:logfmt}, LogFMT enables low-precision token transmission with minimal impact on model performance. Incorporating LogFMT natively within network hardware could further optimize communication by increasing entropy density and reducing bandwidth usage. Hardware-accelerated compression and decompression would allow seamless integration of LogFMT into distributed systems, enhancing overall throughput.

\subsection{Memory-Centric Innovations}
\subsubsection{Limitations of Memory Bandwidth} 
The exponential growth in model sizes has outpaced advancements in high-bandwidth memory (HBM) technology. This disparity creates a memory bottleneck, particularly in attention-heavy architectures like Transformers.

\vspace{-0.4em}
\subsubsection{Suggestions:}
\vspace{-0.2em}
\begin{itemize}[leftmargin=10pt]
    \item \textbf{DRAM-Stacked Accelerators:} Leveraging advanced 3D stacking technologies, DRAM dies can be vertically integrated atop a logic die, thereby enabling exceptionally high memory bandwidth, ultra-low latency, and a practical memory capacity (though stack-limited). This architectural paradigm proves remarkably advantageous for ultra-fast inference in MoE models, where memory throughput is a critical bottleneck. Architectures such as SeDRAM\citep{10185427} exemplify the potential of this approach, delivering unprecedented performance for memory-bound workloads.
    \item \textbf{System-on-Wafer (SoW):} Wafer-scale integration~\cite{cerebras} can maximize computational density and memory bandwidth, addressing the needs of ultra-large-scale models.
\end{itemize}

\section{Conclusion}
\label{sec:conclusion}
DeepSeek-V3 exemplifies the transformative potential of hardware-software co-design in advancing the scalability, efficiency, and robustness of large-scale AI systems. By addressing the limitations of current hardware architectures and proposing actionable recommendations, this paper provides a roadmap for the next generation of AI-optimized hardware. These innovations will be critical as AI workloads continue to grow in complexity and scale, driving the future of intelligent systems.

\bibliographystyle{ACM-Reference-Format}
\bibliography{sample-base, ref_1}

\clearpage 

\end{document}